\documentclass[aps,prl,twocolumn,nofootinbib]{revtex4-1}
\usepackage[a4paper,top=2.5cm,bottom=2.5cm,left=2.0cm,right=2.0cm]{geometry}
\usepackage{amssymb}
\usepackage{amstext}
\usepackage{amsmath}
\usepackage{graphicx}
\usepackage{float}
\usepackage[pdftex]{hyperref}
\usepackage{mathptmx}
\renewcommand{\vr}{{\mathbf r}}
\newcommand{\vz}{{\mathbf z}}

\newcommand{\vp}{{\mathbf p}}
\newcommand{\upt}{UPt$_{3}$}
\newcommand{\sro}{Sr$_{2}$RuO$_{4}$}
\newcommand{\He}{$^{3}$He}
\newcommand{\Hea}{$^{3}$He-A}
\newcommand{\Heb}{$^{3}$He-B}

\newcommand{\SO}{\mathsf{SO}}

\newcommand{\nm}{\,\mbox{nm}}
\newcommand{\mbar}{\,\mbox{bar}}
\newcommand{\angstrom}{\textup{\AA}}
\newcommand{\Pz}{$\text{P}_\mathbf{z}$}
\newcommand{\Bi}{$\text{B}_\mathbf{SO(2)}$}
\newcommand{\Ai}{$\text{A}_\mathbf{SO(2)}$}
\newcommand{\At}{$\text{A}_\mathbf{C_2}$}
\newcommand{\As}{$\text{S}_\mathbf{A}$}
\newcommand{\Bs}{$\text{S}_\mathbf{B}$}
\newcommand{\wpre}{\frac{\gamma^{2}}{\chi_{N}}g_{D}}

\newcommand{\Drr}{\left\langle \Delta_r^2 \right\rangle}
\newcommand{\Dtt}{\left\langle \Delta_\phi^2 \right\rangle}
\newcommand{\Dzz}{\left\langle \Delta_z^2 \right\rangle}

\newcommand{\betawc}[1]{\beta_{#1}^{\text{wc}}}
\newcommand{\betasc}[1]{\beta_{#1}^{\text{sc}}}

\newcommand{\rez}{\Delta_{z}^{\prime}}
\newcommand{\imz}{\delta_{z}^{\prime\prime}}

\newcommand{\imr}{\Delta_{r}^{\prime\prime}}
\def\nicefrac#1#2{\genfrac{}{}{}{1}{#1}{#2}}
\def\point#1#2{{\mathsf{#1}}_{\mbox{\footnotesize #2}}}
\def\ns{\negthickspace}
\def\ms{\negmedspace}
\begin{document}
\title{Spontaneous helical order of a chiral p-wave superfluid confined in nano-scale channels}
\author{J. J. Wiman}
\author{J. A. Sauls}
\email{sauls@northwestern.edu}
\affiliation{Department of Physics and Astronomy, Northwestern University, Evanston, IL 60208}
\date{\today}
\begin{abstract}
Strong interactions that favor chiral p-wave pairing, combined with strong 
pair breaking by confining boundaries, are shown to lead to new equilibrium states 
with different broken symmetries.
Based on a strong-coupling extension of the Ginzburg-Landau (GL) theory that 
accurately accounts for the 
thermodynamics and phase diagram of the bulk phases of superfluid \He{}, we predict new 
phases of superfluid \He{} for confined geometries that spontaneously break rotational and 
translational symmetry in combination with parity and time-reversal symmetry.
One of the newly predicted phases exhibits a unique combination of chiral and helical 
order that is energetically stable in cylindrical channels of radius approaching the Cooper 
pair coherence length, e.g. $R\sim 100\,\mbox{nm}$.
Precise numerical mimimization of the free energy yields a broad region of stability 
of the helical phase as a function of pressure and temperature, in addition to three 
translationally invariant phases with distinct broken spin- and orbital rotation symmetries.
The helical phase is stable at both high and low pressures and favored by 
boundaries with strong pair-breaking.
We present calculations of transverse NMR frequency shifts as functions of \emph{rf} 
pulse tipping angle, magnetic field orientation, and temperature as signatures of 
these broken symmetry phases.
\end{abstract}
\maketitle
{\it Introduction} --- The superfluid phases of \He{} are paradigms for spontaneous 
symmetry breaking in condensed matter and quantum field theory \cite{vol13,sau17}.
The bulk A- and B phases are BCS condensates of p-wave, spin-triplet Cooper pairs \cite{leg75}.
The broken symmetries of these phases, which are well established,
underpin the non-trival topologies of both ground states \cite{volovik03,miz16}.
However, the bulk phases are only \emph{two realizations} of the 18-dimensional manifold 
of spin-triplet, p-wave condensates.
When \He\ is subjected to a confining potential on scales approaching the Cooper pair 
coherence length, $\xi_0 \approx 160-770 \,\angstrom{}$ depending on pressure, new ground 
states with novel broken symmetries are stabilized \cite{bar75,vor07,wim15,wim16}.

In this Letter we report theoretical predictions of the equilibrium phases of superfluid \He\ 
when confined in quasi-one-dimensional channels with radial confinement ranging from
$R = 2-20\,\xi_{0}(p)$.
Among these phases is a novel ``helical'' phase of \He{} that spontaneously breaks both 
time-reversal and translational symmetry along the channel. The broken translational symmetry
is realized as a \emph{double helix} of disclination lines of the chiral axis confined on 
the boundary of the cylinder walls. The double-helix phase is predicted to be stable over a 
large region of the pressure-temperature phase diagram for channels with radius 
$R=100\,\mbox{nm}$.

{\it Ginzburg-Landau Theory} --- Our results are based on a strong-coupling extension of 
Ginzburg-Landau (GL) theory that accurately reproduces the relative stability fo the bulk 
A- and B-phases, including the A-B phase transition \cite{wim15}.
The GL theory is formulated as a functional of the order parameter, the
condensate amplitude for Cooper pairs,
$\langle \psi_{\sigma}(\vp)\psi_{\sigma'}(-\vp)\rangle$ in the spin-momentum basis.
For spin-triplet, p-wave Cooper pairs the order paramater can be expressed in terms of 
a $3\times3$ matrix $A_{\alpha i}$ of complex amplitudes that transforms as the vector
representation of $\point{SO(3)}{S}$ with respect to the spin index $\alpha$, and 
as the vector representation of $\point{SO(3)}{L}$ with respect to the orbital
momentum index $i$.
In cylindrical coordinates the order parameter matrix may be represented as
\begin{equation}
A = 
\begin{pmatrix} 
A_{rr} 		& A_{r \phi} 		& A_{rz} 	\\
A_{\phi r} 	& A_{\phi \phi} 	& A_{\phi z} 	\\
A_{zr} 		& A_{z\phi} 		& A_{zz} 
\end{pmatrix}
\,,
\end{equation}
where we choose aligned spin and orbital coordinate axes. 
The GL free energy functional,
\begin{equation}
\Omega[A] = \int_{V} d^3r\,\left( f_{\mathrm{bulk}}[A] + f_{\mathrm{grad}}[A] \right) 
\,,
\end{equation}
is expressed in terms of a bulk free energy density \cite{thu87},
\begin{align} 
f_\mathrm{bulk}[A] 
&=  
\alpha(T) Tr\left(A A^{\dagger}\right) 
+\beta_{1} \left|Tr(A A^{T})\right|^{2}
\nonumber \\
&
+\beta_{2} \left[Tr(A A^{\dagger})\right]^{2}  
+\beta_{3}\, Tr\left[A A^{T} (A A^{T})^{*}\right] 
\nonumber \\
&
+\beta_{4}\, Tr\left[(A A^{\dagger})^{2}\right]
+\beta_{5}\, Tr\left[A A^{\dagger} (A A^{\dagger})^{*}\right] 
\,,
\end{align}
and the gradient energies,
\begin{equation}
f_\mathrm{grad}[A] 
=K_{1} A_{\alpha j , k}^{*} A_{\alpha j , k} 
+K_{2} A_{\alpha j , j}^{*} A_{\alpha k , k} 
+K_{3} A_{\alpha j , k}^{*} A_{\alpha k , j}
\,,
\end{equation}
where $A^{\dag}$ ($A^{T}$) is the adjoint (transpose) of $A$, 
$A_{\alpha i,j}=\partial_j A_{\alpha i}$, and the transformation
of the gradient free energy from the Cartesian representation to cylindrical coordinates 
given in Eq. (6) of Ref. \onlinecite{wim15}.
The material parameters, $\alpha$, $\{\beta_i |\, i=1\ldots 5\}$, and $\{K_a|\, a=1,2,3\}$
multiplying the invariants defining the GL functional are determined by the
microscopic pairing theory for \He. In weak-coupling theory these parameters are 
given in Refs. \onlinecite{thu87,wim15}.

Ginzburg-Landau theory is widely used in studying inhomogeneous superconducting phases, notably vortex states in type II superconductors \cite{abr57}, as well as Fulde-Ferrell-Larkin-Ovchinnikov states at high field and low temperatures \cite{agt01}.
In the case of \He\ 
a strong-coupling extension of the weak-coupling GL theory that accounts for the relative stability of the bulk A- and B phases, and specifically the A-B transition line, $T_{\text{AB}}(p)$ for pressures above the polycritical point, $p\gtrsim p_c$ was introduced in Ref. \onlinecite{wim15}. 
The strong-coupling functional is defined by the corrections to the fourth-order weak-coupling material parameters, 
\begin{equation}
\beta_i(p,T)=\betawc{i}(p,T_c(p))+\nicefrac{T}{T_c}\Delta\betasc{i}(p)
\,,
\end{equation}
with $\Delta\betasc{i}(p)=\beta_i(p,T_c(p))-\betawc{i}(p,T_c(p))$. 
The weak-coupling parameters, $\beta_i^{\text{wc}}(p,T_c(p))$ are calculated from a 
Luttinger-Ward formulation of the weak-coupling microscopic free-energy functional and 
evaluated using the known pressure-dependent Fermi-liquid material parameters, provided in 
Table I of Ref. \onlinecite{wim15}.
The $\Delta\beta_i^{\text{sc}}(p)$ have been obtained from analysis of measurements of the 
strong-coupling enhancement of heat capacity jumps, NMR frequency shifts and the Zeeman 
splitting of superfluid transition in a magnetic field \cite{cho07}.
The results we report are based on the strong-coupling parameters reported in Table I of Ref. \cite{wim15}.
We emphasize that the extended GL functional accounts for the relative stability of competing phases at temperatures well below $T_c(p)$, including the bulk A and B phases at high pressures \cite{wim15}, and the A to stripe phase transition in thin films of \He\ \cite{wim16}, and in the former case has been validated by our microscopic calculations of $T_{\text{AB}}(p)$ and the strong-coupling beta parameters, $\Delta\beta^{\text{sc}}_i(p)$, based on the formulation of the strong-coupling theory developed in Refs. \onlinecite{rai76,sau81b,sau81c,ser83}.
 
\begin{figure}[t] 
\begin{center}
\includegraphics[width=\columnwidth]{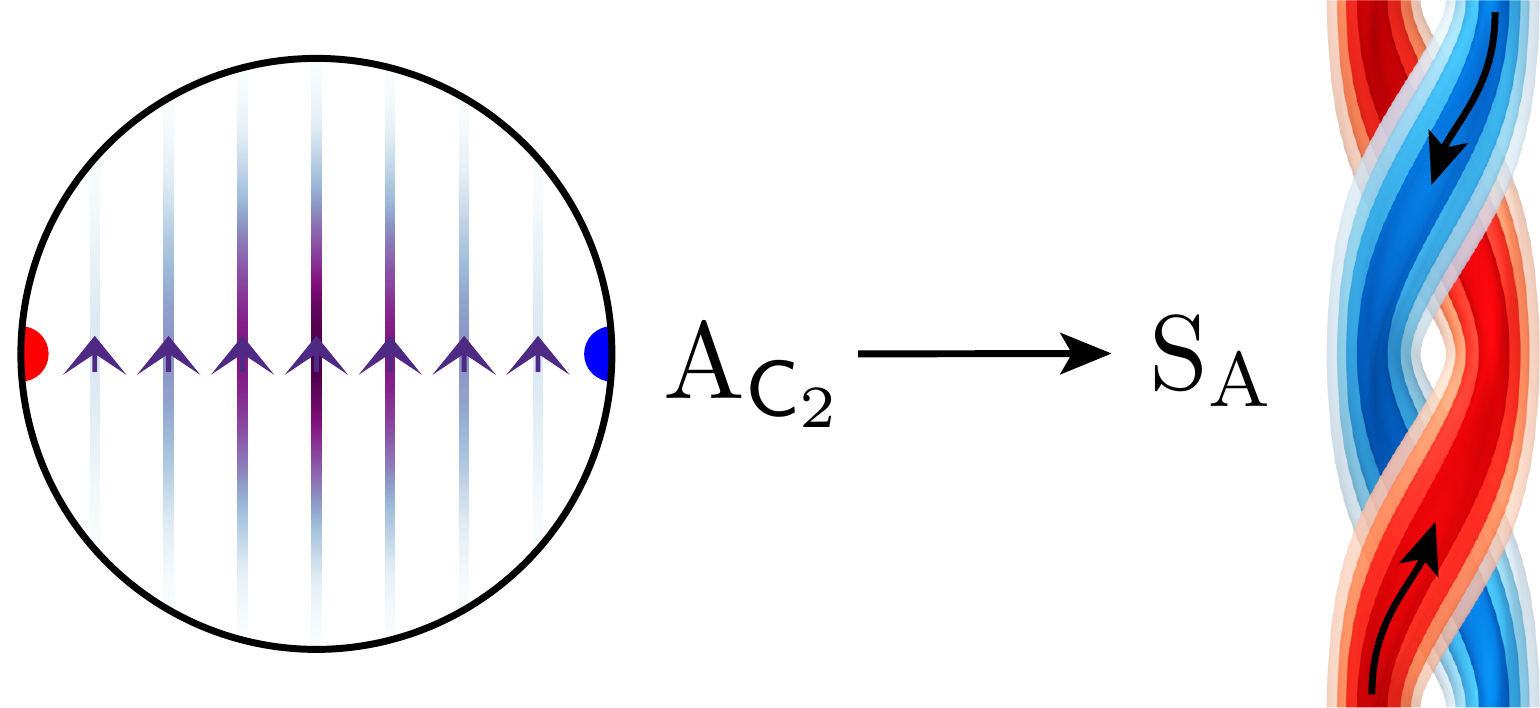}
\caption{Left: The chiral axis $\hat{l}(\vr)$ for the \At{} phase at $p=26\,\mbox{bar}$, 
$T=0.7\,T_c$ with 
strong pairbreaking ($b_T^{\prime}=0.1$). The chiral axis is confined in the $r-\phi$ plane.
The arrow color density is scaled by the amplitude, $(\Delta_r^2+\Delta_\phi^2)^{1/2}$. The red 
and blue dots locate the two disgyrations, which support supercurrents
propagating along $+z$ and $-z$, respectively.
Right: Supercurrent isosurfaces in the \As{} phase, calculated using the full order parameter in Eq. \ref{eq:op-full} of the Appendix.
}
\label{c2-vector}
\end{center}
\end{figure}

The geometry we consider here is an infinitely long cylindrical channel of radius $R$. For the 
channel walls we use boundary conditions that include a variable order parameter ``slip
length'' $b_T$ inspired by the analysis of Ambegaokar, de Gennes, and Rainer \cite{amb74}, as 
well as the influence of boundary curvature \cite{buc77}. The
resulting conditions at $r=R$ are \cite{wim15},
\begin{eqnarray}
A_{\alpha r}\vert_{r=R} = 0
\,,\quad
\frac{\partial A_{\alpha z}}{\partial r}\Big\vert_{r=R} 
&=& 
-\frac{1}{b_T}A_{\alpha z}\vert_{r=R}
\,, 
\nonumber\\
\frac{\partial A_{\alpha \phi}}{\partial r}\Big\vert_{r=R} 
&=& 
\left(\frac{1}{R}-\frac{1}{b_T}\right)A_{\alpha \phi}\vert_{r=R}
\,.
\label{GL-boundary_conditions}
\end{eqnarray}
where the transverse extrapolation parameter $b_T^{\prime}\equiv b_T/\xi_0$ varies between the 
$b_T^{\prime}\rightarrow 0$ (maximal pairbreaking) and $b_T^{\prime}\rightarrow\infty$ (minimal 
pairbreaking) limits. 

The equilibrium order parameter is obtained by minimizing the GL free energy functional, i.e.
by solving the Euler-Lagrange equations, $\delta\Omega[A]/\delta A^\dagger=0$.
When restricted to translationally invariant states we obtain four phases stable in different 
regions of the $p-T$ phase diagram: the \Pz{} phase with Cooper pairs nematically aligned 
along the axis of the cylindrical channel is the first unstable mode from the normal state. 
At a lower temperature Cooper pairs with orbital wave functions transverse to $z$ become 
unstable. 
Strong-coupling and strong pair breaking on the boundary lead to two distinct chiral phases 
with different symmetries. The first is a second-order transition from the \Pz{} phase to the 
\At{} phase with the chiral axis aligned in the plane perpendicular to the $z$ axis.
The \At{} phase spontaneously breaks $\point{SO(2)}{\ns}$ rotation symmetry.  
At lower temperatures the cylindrically isotropic chiral phase, \Ai{}, is stabilized,
and at even lower temperatures, the polar-distorted \Bi{} phase is favored. Both \Ai{} and 
\Bi{} phases are separated by first-order transitions \cite{wim15}.

{\it Dynamical Instability} --- The chiral \At{} phase is an inhomogenous analog of bulk \Hea, 
with a spatially averaged angular momentum (chiral) axis $\langle\hat{l}\rangle$ aligned along
a fixed but arbitrary direction in the $r-\phi$ plane, as shown in Figure \ref{c2-vector}. For 
maximal pairbreaking boundary conditions the \At{} order parameter is given by
\begin{equation}\label{eq:ac2-var}
\hspace*{-2mm}
A_{\alpha i}\ms=\ms\hat{d}_\alpha\cos\left(\frac{\pi r}{2 R}\right)\ms
\left\lbrace 
\rez\hat{z}_i
\ns+\ns
i\,\imr(\cos(\phi\ns-\ns\vartheta) \hat{r}_i\ns-\ns\sin(\phi\ns-\ns\vartheta)\hat{\phi}_i)
\right\rbrace.
\end{equation}
where $\vartheta+\pi/2$ is the angle of the average direction of the angular momentum axis, 
$\langle\vec{l}\rangle$ in the $r-\phi$ plane. The in-plane chiral axis spontaneously breaks 
the continuous $\SO(2)$ rotational symmetry of the confining potential. The corresponding 
continuous degeneracy of the \At{} phase implies the existence of a Nambu-Goldstone (NG) mode 
associated with massless, long-wavelength excitation of the orientation, $\vartheta$,
of $\langle\vec{l}\rangle$.

\begin{figure}[t]
\begin{center}
\includegraphics[width=\columnwidth]{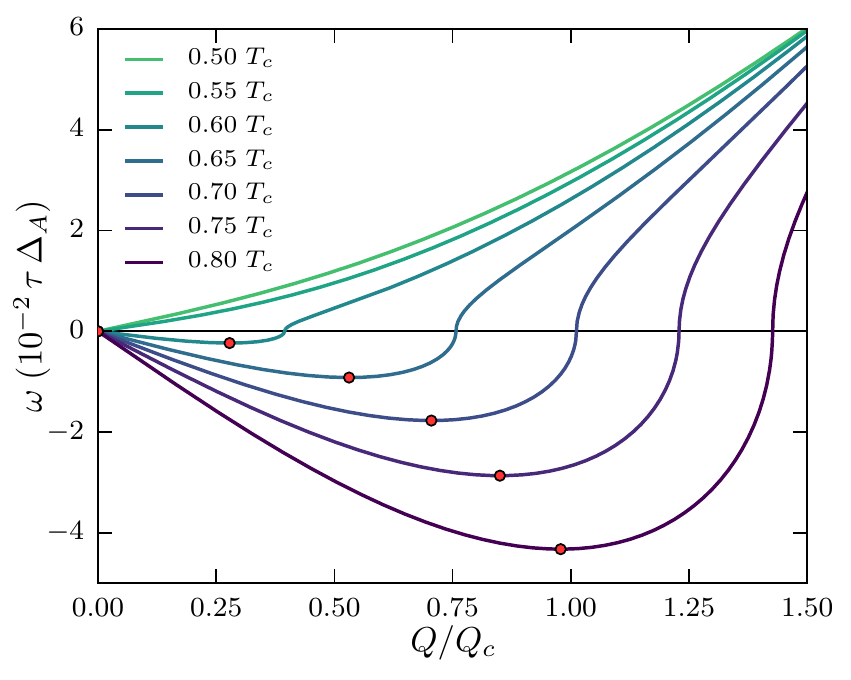}
\caption{NG mode dispersion, $\omega_-$, as a function of $Q$ and $T$ and scaled by the 
bulk A-phase amplitude at $p=26\mbar{}$ and $R=100\nm{}$. Negative values denote imaginary 
values of $\omega_{-}$. The circles indicate the most unstable mode, $Q/Q_c$, for each 
temperature. $Q_c \approx \pi/674\nm{} = 4.66\times 10^{-3}\mbox{nm}^{-1}$ is the 
maximum value of the most unstable mode at the \As{}-\Pz{} transition.}
\label{w-stack}
\end{center}
\end{figure}

The dynamical equation for the NG mode is obtained from the action for the space-time
fluctuations of the Cooper pairs relative to the \At{} ground state,
$\mathcal{A}_{\alpha i}(\vr,t) = A_{\alpha i}(\vr,t)-A_{\alpha i}^{\text{\At}}(\vr)$,
\begin{equation}
\mathsf{S} = 
\int_{V}dt\,d^3r\,\left\lbrace \tau\,\mathrm{Tr}\left(\dot{\mathcal{A}} 
\dot{\mathcal{A}}^{\dagger}\right)  - \mathcal{U}[\mathcal{A}] 
\right\rbrace \,,
\end{equation}
where $\mathcal U[\mathcal A]$ is the effective potential derived from an expansion of 
the free energy functional, $\Omega[A]$, to quadratic order in 
the fluctuations, $\mathcal A$, of order parameter.
The additional invariant represents the kinetic energy of the Cooper pair fluctuations, with
the effective inertia given by $\tau=7\zeta(3)N_f/48(\pi k_{\text{B}} T_c)^2$ in the 
weak-coupling BCS limit \cite{miz18c}, where $N_f$ is the normal-state density of states 
at the Fermi energy.

For the NG mode the action is a functional of the degeneracy variable corresponding to 
space-time fluctuations of the orientation of the chiral axis, $\vartheta(t,z)$, and 
the fluctuations of the polar component of the Cooper amplitude, $\imz(t,z)$, that couples
linearly to $\vartheta(t,z)$ through the gradient energy.
The order parameter that incorporates these fluctuations is
\begin{align}
A_{\alpha i} &= \hat{d}_\alpha \cos\left(\frac{\pi r}{2 R}\right)
\left\lbrace\vphantom{ 
\cos\left(\frac{\pi r}{2 R}\right)}
i\,\imr  \cos[\phi-\vartheta(t,z)] \hat{r}_i \right. \nonumber \\
&\qquad\quad \left.
-i\,\imr  \sin[\phi-\vartheta(t,z)]\hat{\phi}_i
+\rez  \hat{z}_i \right\rbrace \nonumber \\
&\qquad\quad 
- i\, \imz(t,z) \sin\left(\frac{\pi r}{R}\right) \sin[\phi-\vartheta(t,z)] 
\hat{d}_\alpha\hat{z}_i
\,,
\label{eq:spiral-var}
\end{align}
where $\imr$ and $\rez$ take their equilibrium values found by minimizing the free energy 
functional with the order parameter in Eq. \ref{eq:ac2-var}. Since the fluctuations depend 
only on time, $t$, and the coordinate, $z$, along the channel, we can integrate out the 
dependences on $r$ and $\phi$. We then express the action in Fourier space, in which case
we obtain a sum over independent Fourier modes of the form,
$\vartheta(t,z)=\vartheta\cos(\omega t + Q z)$ and $\imz(t,z)=\imz\sin(\omega t + Q z)$. 
The Euler-Lagrange equations reduce to eigenvalue equations for the coupled mode amplitudes,
\begin{align}
\omega^2\,\vartheta &= c^2 Q^2\,\vartheta 
		     + \frac{8(3\pi-4)}{9(\pi^2-4)\imr}\,\left(\frac{\pi c}{R}\right)\,cQ\,\imz
\\
\omega^2\,\imz      
		    &= \frac{1}{\tau}\left\lbrace \alpha
		     + \left(1-\frac{16}{9\pi^2}\right)(\beta_{13}+\beta_{245})
		     (\imr{}^2+\rez{}^2)\right.
\nonumber \\
		    &\quad\; \left.
		     + 3\,c^2 Q^2 
		     + \left(\frac{\pi c}{R}\right)^2
		       \left[1 + \frac{2}{\pi^2}\,\mathrm{Cin}(2\pi)\right]
		       \right\rbrace
		       \imz
\nonumber \\
		    &+ \frac{16(3\pi-4)\imr}{9\pi^2}\,\left(\frac{\pi c}{R}\right)\,cQ\,\vartheta
\,,
\end{align}
where $\mathrm{Cin}(2\pi)=\int_{0}^{2\pi} du \, (1-\cos{u})/u$. The weak-coupling relation 
$K_1=K_2=K_3\equiv K$ has been used, and we introduced the velocity, 
$c\equiv\sqrt{K/\tau}=v_f/\sqrt{5}$, where $v_f$ is the Fermi velocity.

\begin{figure}[t]
\begin{center}
\includegraphics[width=\columnwidth]{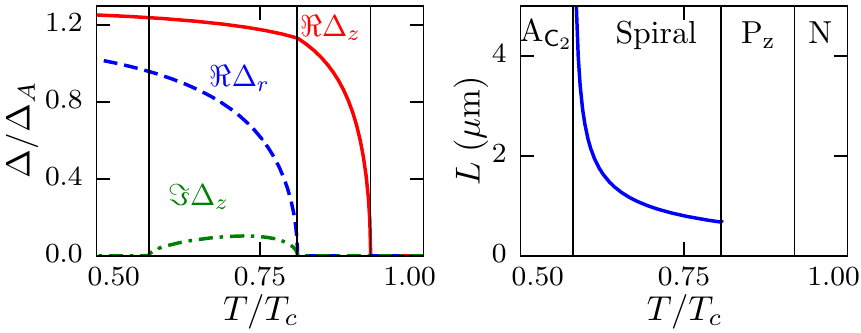}
\caption{(Left) The amplitudes $\rez$, (red); $\imr$, (blue); and $\Delta_z^{\prime\prime}$, 
(green) for the \As{} order parameter phase, at $p=26\mbar{}$, $R=100\mbox{nm}$ and scaled 
by the bulk A phase amplitude $\Delta_A=\sqrt{|\alpha(T)|/4\beta_{245}}$. The black vertical 
lines denote the continuous phase transitions \At{}-\As{}, \As{}-\Pz{}, and \Pz{}-Normal with 
increasing temperature. (Right) The temperature dependence of the half-period $L$.}
\label{spiral-variational}
\end{center}
\end{figure}

There are two eigenmodes corresponding to bosonic excitatoins with dispersions $\omega_{\pm}(Q)$.
The low frequency mode, $\omega_{-}(Q)$ is identified as the NG mode with an excitation 
that is a pure rotation by $\vartheta$, with a linear dispersion $\omega_{-}(Q) \propto Q$
for $Q\rightarrow 0$. 
Indeed the \At{} phase supports low-frequency bosonic excitations corrsponding to oscillations
of the chiral axis, as shown in Fig. \ref{w-stack} for $R=100\,\mbox{nm}$, $p=26\,\mbox{bar}$ 
and $T/T_c = 0.5$. However, the mode softens as the temperature increases.
Above a critical temperature of $T^*\approx 0.57 T_c$ the stiffness supporting the NG mode 
vanishes, and a conjugate pair of imaginary eigenfrequencies appear signalling a helical 
instability of the \At{} phase. 
Figure \ref{w-stack} shows the evolution from the dispersion relation from the region of a 
stable \At{} phase indicated by positive frequencies. Negative values correspond to the 
magnitude of the imaginary frequencies of the unstable NG mode. 
The wavevector of the most unstable mode, $Q_c(T)$, is indicated at each temperature.
As we show below the instability is stabilized to a new chiral phase with spontaneously 
broken translation symmetry along $z$ by nonlinear terms in the GL free energy.

{\it Double Helix Phase} --- The structure of the broken translation symmetry of this new 
phase, designated as \As{}, is that of a \emph{double helix}, easily visualized by the 
propagating rotation of the pair disgyrations as shown in Fig. \ref{c2-vector}.
This phase has continuous \emph{helical} symmetry under the set of rotations 
by $-\Theta$ about $\hat\vz$, $R_z[-\Theta]$ combined with the translation 
along $z$ a distance $+\Theta/Q$, $T_z[+\Theta/Q]$. 
Note also the helical flow of the counter-propagating supercurrents that are confined
near the two disgyrations.
The model for the order parameter in Eq. \ref{eq:spiral-var} allows us to to study the 
temperature evolution of the equilibrium \As{} phase, with rotary propagation
$\vartheta(z)=Qz$, shown in Fig. \ref{spiral-variational}.
Note that half-period, $L=\pi/Q$, is a minimum at the \As{}-\Pz{} transition, 
with $L\approx 37\,\xi_0 \approx 674\, \mbox{nm}$ at $p=26\mbar{}$, 
and diverges as the \As{}-\At{} transition is approached.
The structure of the \As{} phase obtained from the variational model, as well as 
the second-order phase transitions between \At-\As{} phases, and \As-\Pz{} phase, 
agree closely with the numerical minimization of the full GL functional (see Appendix).

{\it Phase Diagram} --- We find \emph{six} distinct phases for cylindrical channels: the 
translationally invariant \Pz{}, \Ai{}, \At{}, and \Bi{} phases reported in Ref. \cite{wim15},
the double helix \As{} phase, and a periodic domain-wall B-phase, \Bs{}, predicted by 
Aoyama \cite{aoy14}.
The \Bs{} phase is defined by domain walls separating polar-distored B-like phases along 
the $z$ axis. We impose boundary conditions for the half-period, $L$, of the order parameter 
at $z=0$ and $z=L$, where $L$ is determined in the minimization of the free energy functional.

\begin{figure}[t]
\begin{center}
\includegraphics[width=\columnwidth]{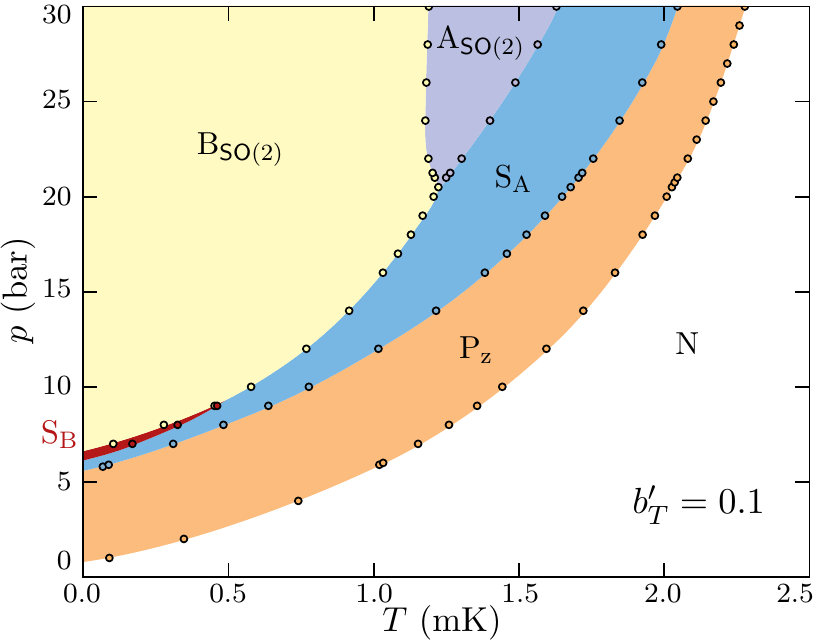}
\caption{Phase diagram for the cylindrical channel with 
$R=100\mbox{nm}$ and strong pairbreaking, $b_T^\prime=0.1$. The labels \As{} 
and \Bs{} correspond to the helical and B-like stripe phases, respectively. 
The \Ai{} phase appears at the highest pressures, and the \At{} phase is suppressed 
by the more stable \As{} phase. 
The \Bs{} phase appears in a narrow region at low pressure and low temperature.}
\label{diagram-strong}
\end{center}
\end{figure}

Figure \ref{diagram-strong} shows the phase diagram for a $R=100\,\mbox{nm}$
cylindrical channel with strong pairbreaking, $b_T^{\prime}=0.1$.
The polar \Pz{} phase, with Cooper pairs  nematically oriented along the channel
is the first superfluid phase to nucleate from the normal state. 
At a lower temperature the transverse orbital components appear; 
the chiral \As{} phase develops at second-order instability
from the \Pz{} phase. Compared to an earlier calculation \cite{wim15}
that assumed translational invariance along the channel, we find that 
the \At{} phase is replaced by the more stable \As{} phase.
At the higher pressures, the isotropic chiral A phase is favored over the helical
phase, separated by a first-order transition line, which then terminates at 
a tricritical point, below which the helical phase is unstable to the polar-distorted 
B phase, \Bi{}, also separated by a first-order transition.
At still lower pressures the \Bs{} phase is stable in a very small window of
the phase diagram.
The \Ai{} and \Bs{} phases are very sensitive to surface pair-breaking, and are completely 
suppressed for maximal pairbreaking (see Appendix).
Finally, as the surface boundary condition approachs specular reflection the \At{} and \As{} 
phases are supplanted by the \Ai{} phase.
A more detailed presentation of the phase diagram, including  
a phase diagram as a function of channel radius $R$, is presented in the Appendix.

{\it NMR Signatures} --- Nuclear magnetic resonance (NMR) spectroscopy is a tool for 
identifing inhomogenous phases of superfluid \He\, \cite{dmi10}. The frequency shift of 
the NMR line relative to the Larmor frequency is sensitive to the spin and orbital 
correlations of the order parameter that minimizes the nuclear magnetic dipole energy, 
$\Delta\Omega_D = \int_V d^3r\,g_D \left( |Tr{A}|^2  + Tr{AA^*} \right)$. 
The dipole energy lifts the degeneracy of the equilibrium states with respect to relative 
spin-orbit rotations. Thus, deviations from the minimum dipole energy configuration
lead dipolar torques generated by the spin-triplet Cooper pairs 
that shift the NMR resonance frequency away from the Larmor frequency.
The magnitude of the shift is determined by the dipole coupling, 
$g_{D}=\nicefrac{\chi_N}{2\gamma^2}\Omega_A^2/\Delta_A^2$,
which can be expressed in terms of normal-state spin susceptibility, $\chi_N$, 
and the bulk A-phase longitudinal resonance frequency, $\Omega_A$. 
We follow the analysis described in Ref. \cite{wim15} for the transverse NMR frequency shifts 
of the translationally invariant phases of \He\ confined in nano-pores to calculate the frequency 
shifts of the \As{} phase.
In particular, the spatially averaged dipole energy density for the \As{} phase is
$f_{D} = g_D \langle\Delta^2\rangle_{\text{\As}}
\left(\hat{d}\cdot\hat{z}\right)^{2}$, with 
$\langle\Delta^2\rangle_{\text{\As}} = 2\Dzz-\Dtt-\Drr$
where $\left\langle\Delta_i^2\right\rangle=\int_V d^3r\,\sum_\alpha\left|A_{\alpha i}\right|^2$
and $\hat{d}$ lies in the plane of the channel and perpendicular to the static magnetic field 
axis $\hat{H}$.
This results in a frequency shift of the same form 
as that of the \Pz{} and \Ai{} phases \cite{wim15}, but with amplitude
$\propto\langle\Delta^2\rangle_{\text{\As}}$,
\vspace*{-2mm}
\begin{align}\label{eq:shift-polar}
\hspace*{-5mm}
\omega\Delta\omega\ns =\ns \wpre{}\langle\Delta^2\rangle_{\text{\As}}
\left[\cos\beta-\sin^{2}\theta\left(\frac{5\cos\beta-1}{4}\right)\right]
\,,
\end{align}

\begin{figure}[t]
\begin{center}
\includegraphics[width=\columnwidth]{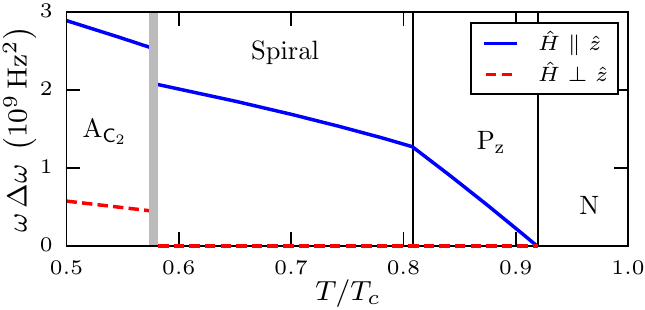}
\caption{Small tipping angle ($\beta\approx0$) transverse frequency shift for 
the \As{} phase as a function of temperature.
The \As{} phase order parameter is that shown in Fig. \ref{spiral-variational}. 
The \As{} and \Pz{} frequency shifts are found using 
Eq. \ref{eq:shift-polar}, and the \At{} frequency shifts are given by Eqs. 36 
and 37 in Ref. \onlinecite{wim15}.
The grey shaded region denotes the region of the \As{} phase 
where the half-period $L$ exceeds $\xi_D\approx 10\;\mu\mathrm{m}$.}
\label{nmrplot}
\end{center}
\end{figure}

\vspace*{-2mm}
\noindent
where $\beta$ is the pulsed NMR tipping angle and $\theta$ is the angle of the static field 
relative to the $z$ axis. 
Figure \ref{nmrplot} shows the frequency shift for $\beta\rightarrow 0$ as a function of 
temperature for the \As{} variational order parameter defined in Eq. \ref{eq:spiral-var} 
and plotted in Fig. \ref{spiral-variational} for two field orientations. The second
order transition at the \Pz-\As{} boundary shows a discontinuity in the slope of 
$\Delta\omega(T)$, and an apparent jump occurs at the \As-\At{} transition.
In fact this is a smooth crossover confined to a narrow temperature range related to the 
divergence of the period of the \As{} phase. The detailed NMR spectrum close to this 
transition is more complex because the spatial variations of the \As{} phase, 
set by the half-period, $L$, can exceed the dipole coherence length, 
$\xi_D\equiv \sqrt{g_D/K_1}\approx 10\,\mu\mbox{m}$ 
near to the \As{}-\At{} transition. The $\hat{d}$ vector becomes inhomogeneous, spatial
averaging breaks down and the NMR line will broaden as the temperature approaches 
the \As-\At{} transition in a narrow window indicated by the gray shading in 
Fig. \ref{nmrplot}.
A narrow NMR line is restored in the \At{} phase.

{\it Conclusions and Beyond \He} --- 
We find six distinct equilibrium phases within highly confined 
cylindrical channels, including two phases that break translation symmetry
along the channel. In particular, we predict a ``helical'' phase, \As{}, which 
spontaneously breaks time-reversal symmetry and translational symmetry, but retains 
rotary-translation (helical) symmetry. The double-helix structure of this phase is 
predicted to be stable over a significant region of p-T phase diagram for long cylindrical 
pores of radius approaching the Cooper pair coherence length $\xi_0$, and to show a distinct
NMR signature.

The novel broken symmetry phases of \He\ are based on competing interactions 
in a strongly correlated Fermi liquid with unconventional pairing, combined with strong 
pair breaking by confining boundaries.
This situation can arise in a broad range of unconventional superconductors, including 
chiral superconductors such as \sro\ and \upt, as well as the cuprates. Indeed theoretical 
predictions of novel broken translational symmetry phases are reported 
for d-wave superconductors subject to strong confinement \cite{vor18,hol18a}, and it seems 
likely that there are more novel broken symmetry phases in multi-component, unconventional 
superconductors awaiting discovery.

{\it Acknowledgements} --- This research was supported by National Science Foundation 
Grant DMR-1508730. We thank Bill Halperin and Andrew Zimmerman for 
discussions on confined phases of \He\ that contributed to this research.
We also thank Michael Moore for bringing Ref. \onlinecite{bar75} to our attention. 

\vspace*{-3mm}
\section{Appendix}
\vspace*{-3mm}

{\it Order Parameter for the Double Helix Phase} ---
Numerical minimization of the GL functional to determine the exact structure of the \As{} 
phase is made efficient, without loss of accuracy, by developing the $\phi$-dependence as 
an expansion in symmetry-preserving harmonics,
\begin{align}
A_{\alpha i} = \hat{d}_\alpha 
\sum_{j=0}^{\infty} 
&\left\lbrace
\vphantom{\sum}\quad \Delta^\prime_{r,j}(r)\sin[2j(\phi+Qz)]\,\hat{r}_i
\right. 
\nonumber\\ 
&+ i\,\Delta^{\prime \prime}_{r,j}(r)\cos[(2j+1)(\phi+Qz)]\,\hat{r}_i 
\vphantom{\sum} 
\nonumber\\
&+ \Delta^\prime_{\phi,j}(r)\cos[2j(\phi+Qz)]\, \hat{\phi}_i 
\vphantom{\sum} 
\nonumber\\
&+ i\,\Delta^{\prime \prime}_{\phi,j}(r)\sin[(2j+1)(\phi+Qz)]\, \hat{\phi}_i 
\vphantom{\sum}
\nonumber\\ 
&+ \Delta^\prime_{z,j}(r)\cos[2j(\phi+Qz)] \,\hat{z}_i 
\vphantom{\sum}
\nonumber\\ 
&\left. \vphantom{\sum}+ i\,\Delta^{\prime \prime}_{z,j}(r)\sin[(2j+1)(\phi+Qz)] \,\hat{z}_i
\right\rbrace 
\,.
\label{eq:op-full}
\end{align}
The numerical result for the \As{} phase converges rapidly to the exact solution with the 
addition of higher harmonics.

{\it Sensitivity of the Phase Diagram to Strong Pairbreaking} --- 
The anisotropic chiral \At{} and \As{} phases are favored under conditions of strong pairbreaking
on  the boundary because the energy cost of the boundary half-disgyrations is minimal due to 
suppression of all the order parameter components. By contrast the \Ai{} phase, which hosts 
a radial disgyration at the center of the cell is disfavored over both anisotropic chiral phases. 
\begin{figure}[t]
\begin{center}
\includegraphics[width=\columnwidth]{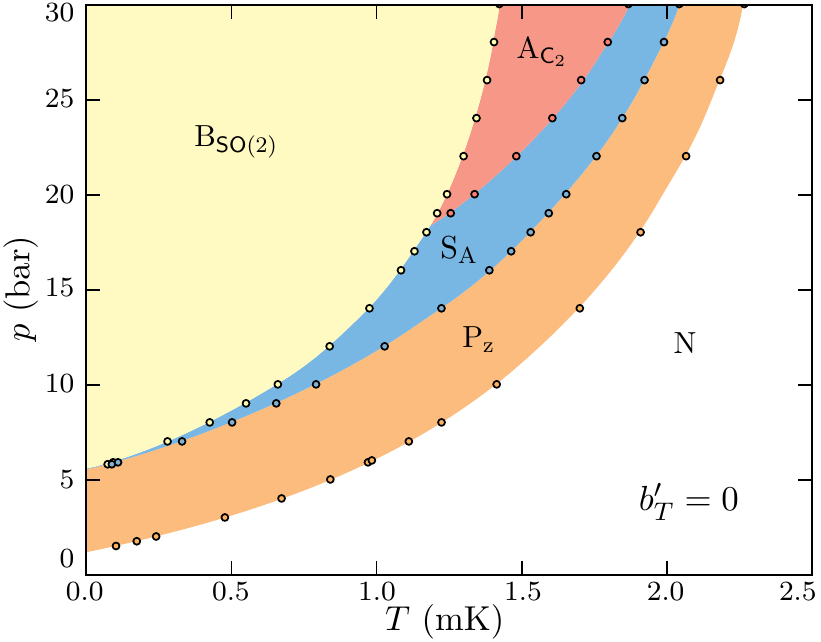}
\caption{Phase diagram for the cylindrical channel with $R=100\mbox{nm}$ 
and $b_T^\prime=0$ (maximal pairbreaking). The \As{} phase has displaced 
much of the \At{} phase compared to the phase diagram calculated for only
$z$ translationally invariant phases \cite{wim15}.
}
\label{diagram-maximal}
\end{center}
\end{figure}

This is reflected in the phase diagram for maximal pairbreaking, $b_T^{\prime}=0$,
shown in Fig. \ref{diagram-maximal} for a $R=100\,\mbox{nm}$
cylindrical channel. Note that strong-coupling, which is relatively stronger at higher 
temperatures favors the helical phase over the translationally invariant \At{} phase.
Also note that \Bs{} phase is not stable at this confinement ($R=100$ nm) for maximal 
pair-breaking.

\begin{figure}[h]
\begin{center}
\includegraphics[width=3.4in]{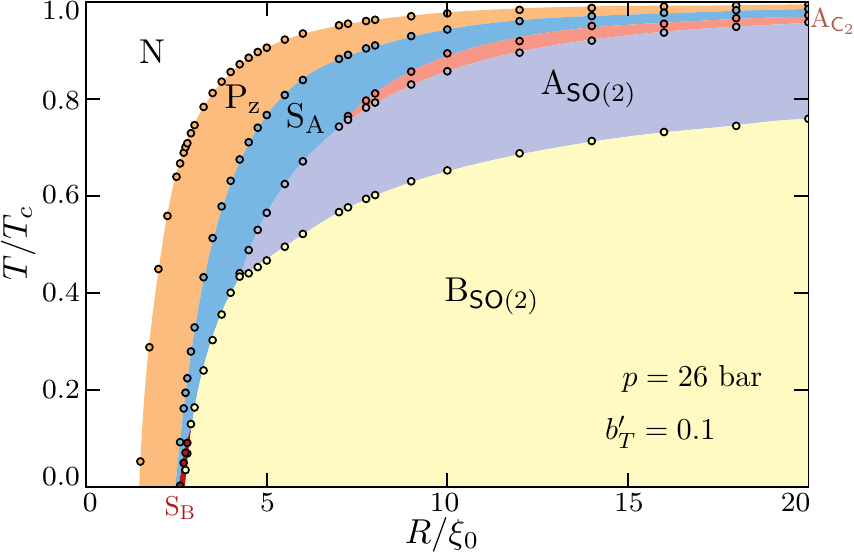}
\caption{Phase diagram: temperature versus radius for cylindrical channel at 
$p=26\mbar{}$ with strong pairbreaking. $b_T^\prime=0.1$. 
All six phases we have found are shown in this diagram, although the \Bs{}
phase is stable in a very narrow window of $R$ and $T$.}
\label{diagram-radius}
\end{center}
\end{figure}

We also include the phase diagram as a function of the channel radius $R$ (Fig.
\ref{diagram-radius}) for $p = 26\mbar{}$ and $b_T^\prime=0.1$. The \As{} 
phase is clearly favored by high confinement relative to the \At{}, \Ai{}, and \Bi{} phases.
However, at this pressure the \Bs{} phase is very fragile and stable only at very low 
temperatures where non-local corrections to the gradient energy, which are not included
in the extended GL functional, are likely relevant.
We also note that there is a critical radius above which the \At{} phase appears.

%
\end{document}